\documentclass{article}
\usepackage{ijcai26}

\usepackage{times}
\usepackage{soul}
\usepackage{url}
\usepackage[hidelinks]{hyperref}
\usepackage[utf8]{inputenc}
\usepackage[small]{caption}
\usepackage{graphicx}
\usepackage{amsmath}
\usepackage{amsthm}
\usepackage{booktabs}
\usepackage{algorithm}
\usepackage{algorithmic}
\usepackage[switch]{lineno}
\usepackage{amssymb}

\title{DySRec: Dynamic Context-Aware Psychometric Scale Recommendation via Multi-Agent Collaboration}

\author{
Yanzeng Li$^{1}$\and
Xiaoning Cao$^{1}$\and
Jialun Zhong$^{2}$\and
Jianpeng Hu$^2$\and
Jiangshan Tan$^{3}$\and \\
Ningning Liu$^{4}$ \and
Feng Xiang$^{5}$ \And
Shasha Han$^{6}$\thanks{Corresponding Author}\\
\affiliations
$^1$Institute of Artificial Intelligence and Future Networks, Beijing Normal University\\
$^2$Wangxuan Institute of Computer Technology, Peking University\\
$^3$Fuwai Hospital, Chinese Academy of Medical Sciences \& Peking Union Medical College\\
$^4$Peking University Sixth Hospital,
Peking University Institute of Mental Health\\
$^5$Beijing Malt Butler Technology Co., Ltd\\
$^6$School of Population Medicine and Public Health, Chinese Academy of Medical Sciences \& Peking Union Medical College
\emails
liyanzeng@bnu.edu.cn, hanshasha@pumc.edu.cn
}

\begin{document}

\maketitle

\begin{abstract}
Choosing suitable psychometric scales is an essential and difficult step in psychological consultation, which requires clinicians to integrate patient information, behaviors, and dynamic contextual information.
Existing systems mainly use static pipelines to choose scale, or directly predict symptoms according to user inputs, limiting their ability to support dynamic assessment, risk management, and transparent decision-making. 
To address these limitations, we propose DySRec, a multi-agent conversational system for dynamic psychometric scale recommendation. 
DySRec operates as an interactive chatbot that engages users in multi-turn dialogue, models scale selection as a continuous conversational decision process, and coordinates specialized agents to maintain user context, recommend assessment scales, monitor psychological risk, and log decision trajectories. 
In this way, DySRec can integrate and capture heterogeneous signals, including semantic, interaction behaviors, assessment history, and content state, to dynamically update user representations and calculate scale-context compatibility score for recommending most matched scales.
Moreover, DySRec incorporates a closed-loop refinement mechanism. Recommendation agent will feedback the missing or uncertain attributes and guide the conversation to elicit the targeted information. 
In this paper, we showcase the prototype design and architecture of DySRec, and this system has been verified in a real-world application.
\end{abstract}

\section{Introduction}

Psychometric assessment is a core component of psychological diagnosis, screening, and intervention planning in both clinical practice and large-scale digital mental health systems~\cite{world2022icd,myers2002ten}. 
Standardized psychometric scales offer structured measures of latent psychological constructs such as depression, anxiety, and cognitive functioning, thereby supporting evidence-based decision-making~\cite{kroenke2001phq,spitzer2006brief}.
However, selecting appropriate scales and adapting assessment strategies over time remains a non-trivial decision problem. 
In practice, clinicians will continuously integrate patient utterances, behavioral signals, prior assessment results, and contextual factors to update their understanding of a patient's psychological state, and determine subsequent assessment steps~\cite{groth2009handbook}. This process is sequential, context-sensitive, and safety-critical~\cite{meyer2001psychological}.

Despite recent advances in artificial intelligence for mental health~\cite{shatte2019machine,lee2021artificial}, most existing automated assessment systems treat psychometric evaluation as a static inference task. These systems typically follow predefined assessment pipelines, or apply supervised models to predict symptom severity, or diagnostic labels from isolated user inputs~\cite{calvo2017natural,hilty2025artificial}. 
Such approaches implicitly assume a fixed assessment policy and do not account for the dynamic evolution of user states or shifting assessment objectives~\cite{liu2019comparison}. 
Recent progress in Large Language Models (LLMs) have highlighted their potential to extract nuanced psychological signals from free-form text~\cite{hua2024large,li2026community}. However, these methods primarily focus on state estimation, e.g., symptom detection or score prediction, rather than adaptive policy planning. Consequently, the challenge of dynamically selecting psychometric scales remains underexplored. This task requires reasoning over evolving belief states, clinical safety constraints, and the costs associated with assessment actions.
To address this gap, there is a need to move beyond monolithic predictive models toward integrated architectures that explicitly represent user state, assessment actions, and the decision-making processes that coordinate them over time.

Moreover, scale selection in real clinical practice is not a purely reactive interview. Clinicians often ask targeted follow-up questions to clarify ambiguous or missing information before administering structured instruments~\cite{dejonckheere2019semistructured,kramer2019introduction}. Inspired by this process, we introduce a feedback-driven refinement mechanism. When critical information is incomplete, the system actively guides the conversation to collect the corresponding attributes to reduce uncertainty for scale recommendation. 

In summary, we propose DySRec, a \underline{dy}namic, psychometric \underline{s}cale \underline{rec}ommendation framework grounded in multi-agent collaboration. DySRec formulates psychometric assessment as a sequential decision-making process, and multiple specialized agents continuously operate and maintain a shared contextual state. Instead of learning a monolithic policy, DySRec decomposes the problem into coordinated roles responsible for belief state maintenance, action recommendation, risk evaluation, execution, and monitoring. This structured decomposition enables explicit modeling of competing objectives, including assessment informativeness, user burden, and safety, while allowing agents to collaborate through a shared context representation. 

\section{Methodology}\label{sec:method}

DySRec models psychometric assessment as a sequential decision process $\mathcal{M} = \langle \mathcal{C}, \mathcal{A}, \mathcal{R}, \mathcal{T} \rangle$, where $\mathcal{C}$ is the evolving context state, $\mathcal{A}$ is the action space of psychometric scales, $\mathcal{R}$ is the risk-aware policy, and $\mathcal{T}$ is the state transition function.

\paragraph{Dynamic Context State Representation.}
The system maintains a global context state $C_t \in \mathbb{R}^d$ that serves as the shared memory for all agents. At each interaction step $t$, the state is updated via a transformation function $\Phi$:
\begin{equation}
    C_t = \Phi(u_t, C_{t-1})
\end{equation}
where $u_t$ is the user's latest utterance and $C_t$ is a concatenated vector representing:
\begin{itemize}
    \item \textbf{Latent Semantic State:} LLM-derived embeddings of emotional valence and symptom mentions.
    \item \textbf{Behavioral Metadata:} Non-verbal signals including response latency and engagement scores.
    \item \textbf{Longitudinal History:} A summarized representation of prior scale scores and diagnostic trajectories.
\end{itemize}

\paragraph{Information-Gain Driven Dialogue Refinement.}
DySRec introduces a closed-loop refinement mechanism to enhance dialogue. Before the recommendation period, the \textit{Recommendation Agent} evaluates the completeness of state via a confidence metric $Conf(C_t)$, defined as the ratio of observed clinical attributes to the required attribute set $\mathcal{K}_{req}$ for the suspected condition:
\begin{equation}
    Conf(C_t) = 1 - \frac{\sum_{i \in \mathcal{K}_{req}} \mathbb{I}(i \notin C_t)}{|\mathcal{K}_{req}|}
\end{equation}
If $\tau_{min} <Conf(C_t) < \tau_{max} $, where $\tau$ are thresholds, the system triggers a \textit{Refinement Loop}. This mechanism is designed to actively guide the user to provide additional information to improve the recommendation accuracy after the preliminary exploratory chat (i.e., $Conf(C_t) > \tau_{min}$). The agent purposefully selects the attribute $a^*$ that maximizes the expected information gain to reduce entropy $H$ in the diagnostic belief state:
\begin{equation}
    a^* = \arg\max_{a} [ H(C_t) - \mathbb{E}[H(C_t | a)] ]
\end{equation}

This design enables active information-seeking behavior and dynamic uncertainty reduction for our system in dialogue period.

\paragraph{Adaptive Scale Scoring and Selection.}
When confidence is sufficient ($Conf(C_t) > \tau_{max}$), candidate scales $s \in \mathcal{S}$ are ranked using a weighted multi-criteria linear scoring function:
\begin{equation}
    S_{cand}(s) = \mathbf{w}^\top \cdot [A(s), Pr(s)]^\top
\end{equation}
The adaptability score $A(s)$ computes the cosine similarity between the current context $C_t$ and the pre-calculated scale's characteristic vector of each scale. The priority score $Pr(s)$ utilizes heuristic weights to prioritize shorter scales (reducing user burden) or complementary instruments that cover different psychological dimensions.

Upon obtaining the candidate set $S_{cand}$ through the aforementioned scoring procedure, the system invokes the LLM to perform discriminative re-ranking and constraint-based filtering, thereby yielding the final selected scale set $S_{final}$ and deliver them to users.

\paragraph{Real-time Risk-Aware Coordination.}
Safety is enforced by a dedicated \textit{Risk Agent} that operates asynchronously. It computes a continuous risk index $R(t)$:
\begin{equation}
    R(t) = \sigma (\alpha E_t + \beta K_t + \gamma L_t + \delta S_t)
\end{equation}
where $E, K, L, S$ represent emotional volatility, high-risk keywords, linguistic anomalies (e.g., pressure speech), and historical severity, respectively. The coefficients $\alpha$,$\beta$,$\gamma$,$\delta$ are weights that determine each factor's contribution to the overall risk score, allowing adaptation to different user populations or contexts. The function $\sigma(\cdot )$ is a sigmoid mapping that bounds $R(t)$ to $[0,1]$, facilitating threshold-based decisions.
If $R(t) > R_{high}$, a hard-constraint override is triggered, bypassing the recommendation logic to initiate immediate clinical intervention protocols.

\section{System Design \& Architecture}

\begin{figure*}[htbp]
\centering{\includegraphics[width=\linewidth]{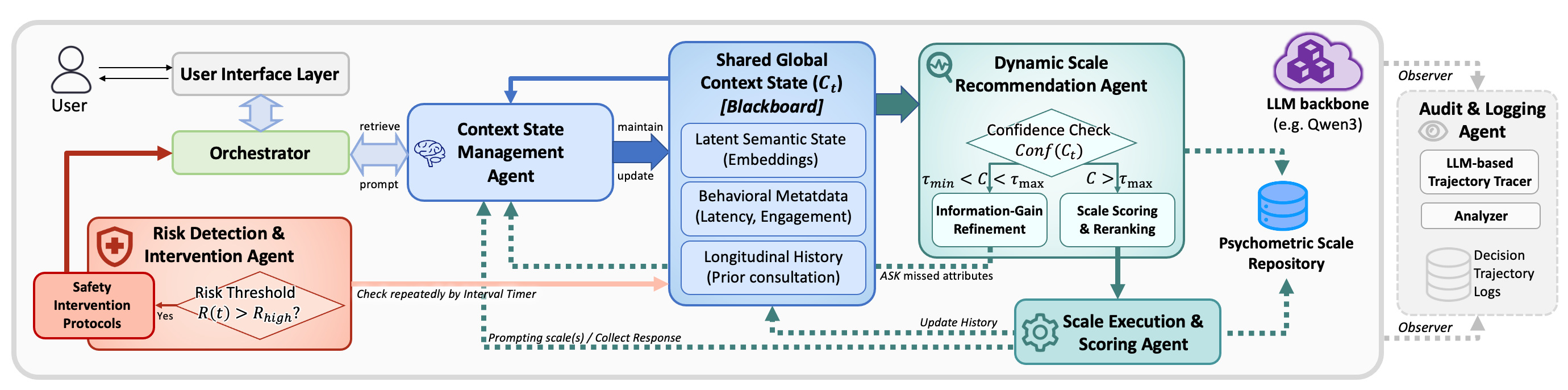}}
        \caption{The overview diagram of DySRec. Different colors and types of arrows represent different data flows, the overall workflow is organized by an Orchestrator, which also serves as a gateway to interact with user interface; communication and information transfer between agents uniformly use JSON with a predefined schema.}
        \label{fig:overall}
    \end{figure*}
DySRec is developed as a modular, multi-agent system that supports dynamic psychometric scale recommendation under evolving user context and safety constraints. Figure~\ref{fig:overall} illustrates the overall architecture.

\subsection{Overall Architecture}

DySRec adopts a modular multi-agent architecture supporting dynamic assessment under evolving context and safety constraints. To ensure seamless coordination, DySRec utilizes a Blackboard-style communication~\cite{han2025exploring} where all agents read from and write to a shared versioned global context state $C$. This allows for asynchronous operations, such as the Risk Agent monitoring the session without blocking others' logic.

\subsection{Agent-level Components}

\paragraph{Context State Management Agent} maintains the shared memory (i.e., the global state $C$), by aggregating multi-dimensional signals, which is described in Section \ref{sec:method}. This agent updates the context state after each user interaction, providing shared and interpretable state for downstream agents. It also processes refinement comes from the Recommendation Agent, generating targeted system prompts and updating the context state accordingly.

\paragraph{Dynamic Scale Recommendation Agent. }
Based on the current context state, the Scale Recommendation Agent performs multi-step decision-making to select appropriate psychometric scales or provide missed attributes/information to Context Agent.
The recommendation progress is detailed in Section~\ref{sec:method}. This agent supports both single-scale and joint multi-scale recommendations to meet the requirements.

\paragraph{Risk Detection \& Intervention Agent.}
To ensure safety in sensitive mental health scenarios, DySRec incorporates this module that asynchronously and continuously estimates the risk index. When predefined risk thresholds are exceeded, the agent can override the recommendation process, halt ongoing assessments, and trigger intervention strategies including forcing the UI to display emergency resources and notifying a human supervisor.

\paragraph{Scale Execution \& Scoring Agent} manages the presentation, progression, and automated scoring of psychometric instruments. Scales are stored in JSON schema-based format, allowing new instruments to be added without modifying system. Assessment results are fed back to the context state to support adaptive follow-up decisions.

\paragraph{Audit \& Logging Agent. }
To support transparency, reproducibility, and regulatory compliance, the Audit Agent records detailed logs of recommendation decisions, risk evaluations, scoring rules, and intervention triggers. These logs enable post-hoc analysis, system debugging, and accountability in real-world deployments.

\begin{figure}[t]
    \centering
    \includegraphics[width=\linewidth]{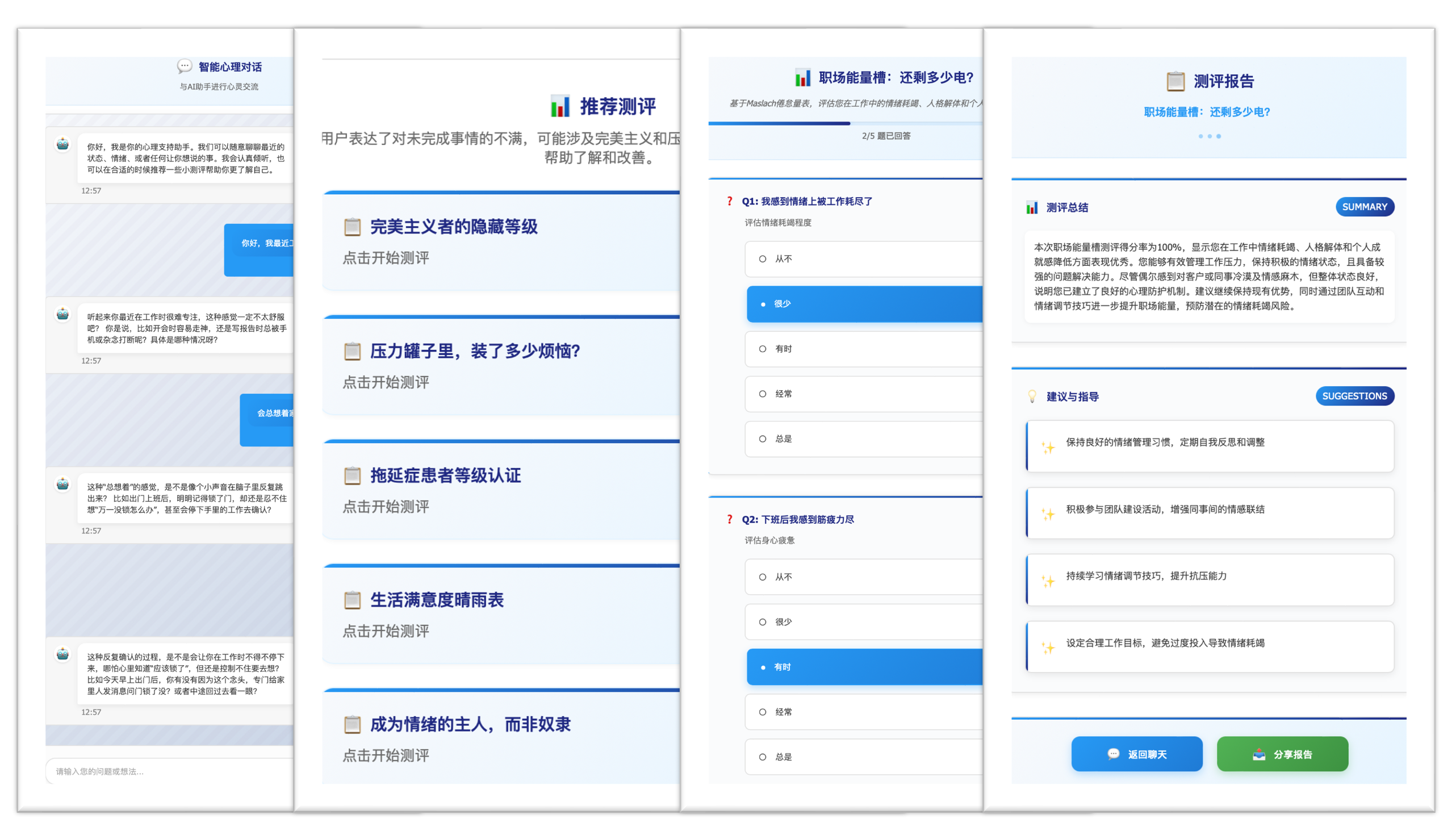}
    \caption{
    Screenshot of DySRec. From left to right, subfigures are: (1) Dynamic dialogue process (2) Scale recommendation results (3) Scale execution (4) Scale scoring and interpretation. }
    \label{fig:screenshot}
\end{figure}

\subsection{Implementation \& Demonstration}

We implemented DySRec using Qwen3~\cite{yang2025qwen3} as the backbone LLM, with thresholds $\tau_{\min}=0.2$, $\tau_{\max}=0.8$, and $R_{\text{high}}=0.85$, determined through pilot calibration with domain experts. The current system prototype has been deployed in a mobile mental health application, called ``Black hole Roast'', serving over 5,000 users, supporting adaptive scale recommendation among 53 diverse psychometric scales. This real-world application has verified the effectiveness of our proposed approach.

\section{Conclusion}

We introduced DySRec, a dynamic context-aware psychometric scale recommendation framework grounded in multi-agent collaboration. By modeling assessment as a sequential decision process, DySRec moves beyond static symptom prediction toward adaptive policy-level reasoning. The system integrates incremental context modeling, adaptive scale scoring, closed-loop information refinement, and safety-driven risk control within an auditable architecture. As a deployed system, DySRec demonstrates how structured agent decomposition and LLM-based semantic modeling can jointly support interactive, transparent, and safety-conscious psychometric workflows\footnote{Demo Video: \url{https://vimeo.com/1165370352}.}.

\bibliographystyle{named}
\bibliography{ijcai26}

\end{document}